\documentclass[letterpaper]{article} % DO NOT CHANGE THIS
\usepackage{aaai25}  % DO NOT CHANGE THIS
\usepackage{times}  % DO NOT CHANGE THIS
\usepackage{helvet}  % DO NOT CHANGE THIS
\usepackage{courier}  % DO NOT CHANGE THIS
\usepackage[hyphens]{url}  % DO NOT CHANGE THIS
\usepackage{graphicx} % DO NOT CHANGE THIS
\usepackage{natbib}  % DO NOT CHANGE THIS AND DO NOT ADD ANY OPTIONS TO IT
\usepackage{caption} % DO NOT CHANGE THIS AND DO NOT ADD ANY OPTIONS TO IT
\usepackage{algorithm}
\usepackage{algorithmic}

\usepackage{amsmath,amsfonts,bm}
\usepackage{graphicx}
\usepackage{multirow}
\usepackage{makecell}
\usepackage{array}
\usepackage[para]{threeparttable}
\usepackage{booktabs}
\usepackage{tabularx}
\usepackage{xspace}
\usepackage{caption}
\usepackage{subcaption}
\usepackage{longtable}
\usepackage{bm}
\usepackage{enumitem}
\usepackage{algorithm}
\usepackage{comment}
\usepackage[colorinlistoftodos]{todonotes}
\usepackage{tablefootnote}
\usepackage{color}
\usepackage{tcolorbox}
\usepackage{tikz}
\usepackage{tabularray}

\usepackage[T1]{fontenc}
\usepackage[utf8]{inputenc}
\usepackage{listings}
\usepackage{xcolor}
\usepackage{xspace}
\usepackage{fontawesome5}

\newcommand{\datawise}{DatasetOracle\xspace}
\newcommand{\oracle}{Oracle\xspace}
\newcommand{\ours}{\textsc{RouterRetriever}\xspace}
\usepackage{newfloat}
\usepackage{listings}
\DeclareCaptionStyle{ruled}{labelfont=normalfont,labelsep=colon,strut=off} % DO NOT CHANGE THIS
\floatstyle{ruled}
\newfloat{listing}{tb}{lst}{}
\floatname{listing}{Listing}
\usepackage{hyperref} %-- This package is specifically forbidden

\title{\ours: Routing over a Mixture of Expert Embedding Models}
\author{
    Hyunji Lee\textsuperscript{\rm 1}\thanks{Work performed during internship at Ai2.} \quad
    Luca Soldaini\textsuperscript{\rm 2} \quad 
    Arman Cohan\textsuperscript{\rm 2, 3} \quad
    Minjoon Seo\textsuperscript{\rm 1} \quad 
    Kyle Lo\textsuperscript{\rm 2}
}
\affiliations{
    \textsuperscript{\rm 1}KAIST AI \quad
    \textsuperscript{\rm 2}Allen Institute for AI \quad
    \textsuperscript{\rm 3}Yale University\\
    \texttt{hyunji.amy.lee@kaist.ac.kr \quad \{lucas, kylel\}@allenai.org}
}

\usepackage{bibentry}
\begin{document}

\maketitle

\begin{abstract}
Information retrieval methods often rely on a single embedding model trained on large, general-domain datasets like MSMARCO. While this approach can produce a retriever with reasonable overall performance, they often underperform models trained on domain-specific data when testing on their respective domains. Prior work in information retrieval has tackled this through multi-task training, but the idea of routing over a mixture of domain-specific expert retrievers remains unexplored despite the popularity of such ideas in language model generation research. 
In this work, we introduce \ours, a retrieval model that leverages a mixture of domain-specific experts by using a routing mechanism to select the most appropriate expert for each query. 
\ours is lightweight and allows easy addition or removal of experts without additional training. 
Evaluation on the BEIR benchmark demonstrates that \ours outperforms both models trained on MSMARCO (+2.1 absolute nDCG@10) and multi-task models (+3.2). This is achieved by employing our routing mechanism, which surpasses other routing techniques (+1.8 on average) commonly used in language modeling. 
Furthermore, the benefit generalizes well to other datasets, even in the absence of a specific expert on the dataset. 
\ours is the first work to demonstrate the advantages of routing over a mixture of domain-specific expert embedding models as an alternative to a single, general-purpose embedding model, especially when retrieving from diverse, specialized domains. 
\end{abstract}

\noindent\textbf{\faGithub\ Code} \hspace{1.2em} \href{https://github.com/amy-hyunji/RouterRetriever}{\footnotesize{\text{\texttt{github/amy-hyunji/RouterRetriever}}}} \\
\noindent\includegraphics[height=1em]{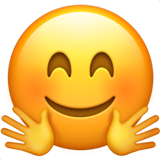} \textbf{Weights} \hspace{0.5em} \href{https://huggingface.co/amy-hyunji-lee/routerretriever}{\footnotesize{\text{\texttt{hf.co/amy-hyunji/RouterRetriever}}}}

\section{Introduction}

Domain-specific retrievers have been shown to outperform general-purpose retrievers for specialized retrieval settings~\citep{izacard2021unsupervised, bonifacio2022inpars}, even in cases where domain-specific training data is only available at a much smaller scale than general-domain datasets like MSMARCO~\citep{Campos2016MSMA}. 
Yet, developing and maintaining separate retrieval systems for each specialized retrieval domain can be costly compared to simply maintaining a single general-purpose MSMARCO-trained model.
Even employing multi-task training, which combines both MSMARCO as well as domain-specific data, to improve performance of the single model setup can be expensive when considering the need for full model retraining whenever a new target retrieval domain emerges, and may not always preserve performance uniformly across all target domains~\citep{wang2023improving, lee2023back}. 
Research has largely focused on improving the performance of these single model setups through data construction~\citep{wang2021gpl, ma2020zero} and domain adaptation~\citep{xin2021zero, Fang2024CombiningMS}.
But less attention has been paid to what can be done if we afford ourselves to use \emph{multiple} specialized retrieval models.

\begin{figure}[t!]
    \centering
        \includegraphics[width=\linewidth]{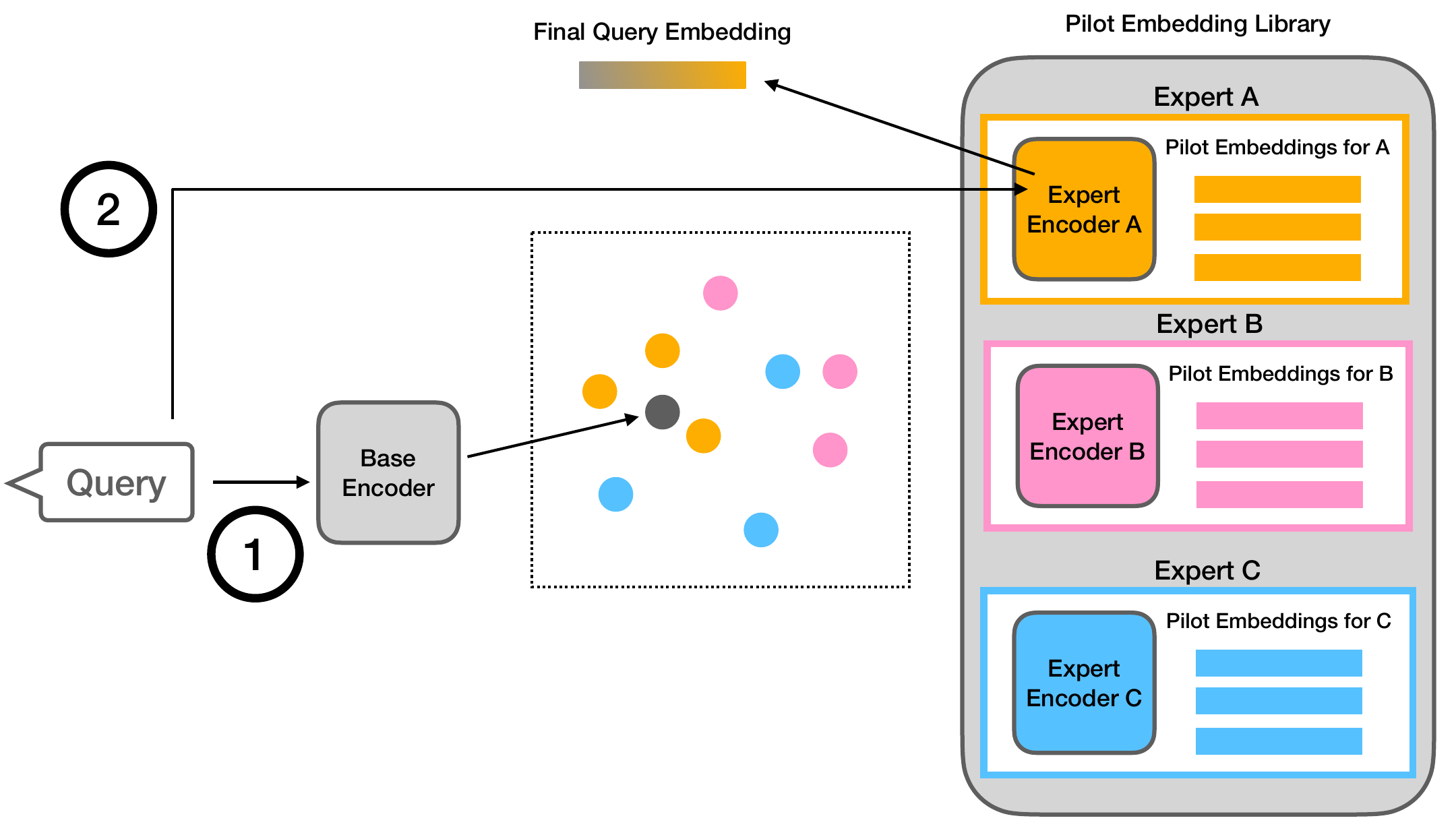}
\caption{\ours: \textcircled{1} Given a query, we first extract its embedding using a base encoder. We then calculate an average similarity between the query embedding (black dot) and the pilot embeddings for each expert (orange dots for Expert A, red dots for Expert B, and blue dots for Expert C). The expert with the highest average similarity (Expert A in this case) is selected. \textcircled{2} The final query embedding is then produced by passing the query to Expert Encoder A, which consists of the base encoder combined with the selected expert LoRA module.} 
\label{fig:ours}
\end{figure}

In this work, we introduce \ours, a retrieval model that leverages a mixture of domain-specific experts with a routing mechanism to select the most suitable expert for each instance. 
\ours consists of a shared base retrieval model and multiple LoRA~\citep{Hu2021LoRALA} components which serves as experts for specific domains.
During training, each expert is trained on a domain-specific dataset while sharing the same frozen base model. 
Thus the expert component captures domain-specific knowledge and extracts embeddings tailored to the domain.
At inference time, as shown in Figure~\ref{fig:ours}, our routing method determines the most appropriate expert by calculating the average similarity between the query and a set of pilot embeddings representing each expert. 
The expert with the highest similarity score is selected, and the corresponding domain-specific embedding is generated using the chosen expert.
\ours is lightweight, as it only requires the training of a parameter-efficient LoRA module for each expert, resulting in a minimal increase in parameters. Additionally, \ours offers significant flexibility: unlike maintaining a single model that requires retraining when domains are added or removed, \ours simply adds or removes experts without the need for further training of the rest of the model.

We demonstrate the effectiveness of \ours on the BEIR benchmark~\citep{thakur2021beir} through a series of experiments with various combinations of target domains:
First, \ours routing between only domain-specific experts outperforms both training a single model on the same dataset in a multi-task manner and training a single model on MSMARCO only. 
Second, we observe that routing techniques from language modeling research don't necessarily translate to our retrieval setting, which motivates us to introduce a new, specialized routing technique based on embedding similarities.
Third, \ours consistently improves performance as we add new experts, whereas multi-task training tends to show performance degradation after a certain number of target domains are included. 
In fact, even without an expert defined for a target domain, \ours can outperform a single general-purpose model on those unseen domains.

We also conduct analysis to better understand the factors behind these performance benefits.
First, to understand data scaling benefits when training each individual expert, we find in-domain performance rapidly increases as we increase training set size, whereas out-of-domain performance often may not see any improvement with scale; hence, the need to use multiple experts.
Second, we show that \ours continually improves as the number of experts increases but quickly runs into diminishing returns.
These improvements aren't just with respect to overall performance, but also with respect to performance stability across domains.
Third, when analyzing routing errors, \ours tends to have a sparser expert selection compared to our instance-level oracle setup. 

\section{Related Works}

\paragraph{Domain Specific Retriever}
Substantial research on retrieval models aims to improve performance on domain-specific tasks.
One approach focuses on dataset augmentation. As domain-specific training datasets are often unavailable and can be costly to construct, researchers have developed methods that either train models in an unsupervised manner~\citep{lee-etal-2019-latent, gao-etal-2021-simcse, gao-callan-2021-condenser} or fine-tune models on pseudo-queries generated for domain-specific datasets~\citep{bonifacio2022inpars, ma2020zero, wang2021gpl}.
Another approach is developing domain-specific embeddings. A common approach is training in a multi-task manner over domain-specific datasets~\citep{Lin2023HowTT, wang2021gpl}. 
Recent works have aimed to improve domain-specific retrievers by developing instruction-following retrieval models~\citep{asai2022task, weller2024followir, oh2024instructir, su2022one, wang2023improving}; instruction contains such domain knowledge.
Another example is \citet{Fang2024CombiningMS} which trains a soft token for domain-specific knowledge. 
While these methods also aim to generate high-quality domain-specific embeddings, they focus on incorporating domain-specific knowledge into the input and processing it with a \emph{single} embedding model.
In contrast, \ours employs a mixture of \emph{multiple} embedding models, encoding domain knowledge directly into their parametric representations to produce more effective embeddings.

\paragraph{Routing Techniques}

Various works have focused on developing domain-specific experts and routing mechanisms to improve general performance in generation tasks.
One approach simultaneously trains experts and the routing mechanism~\citep{sukhbaatar2024branch, muqeeth2024learning}.
Another line of work includes post-hoc techniques that do not require additional training for routing.
Some approaches use the model itself as the knowledge source by training it on domain-specific knowledge~\citep{feng2023knowledge}, incorporate domain-specific knowledge in the token space~\citep{belofsky2023token, shen2024learning}, or select the most relevant source from a sampled training dataset of each domain~\citep{ye2022retrieval, jang2023exploring}. 
Routing techniques have also been investigated for improving generation quality in retrieval-augmented generation tasks; \citet{mallen2022not} explores routing to decide whether to use external knowledge and \citet{jeong2024adaptive} focuses on routing to choose among different retrieval approaches.
In our work, we observe that directly adapting routing techniques from generation tasks to retrieval does not yield optimal performance. 
To address this, we introduce a routing technique specifically tailored to retrieval tasks.
In information retrieval, \cite{Lin2023DecomposingCQ} introduces a technique that decomposes long and complex queries into sub-queries, which are then routed to specialized expert retrievers. Unlike our work, which employs lightweight components as experts, they rely on separate, individual expert models. 
Further, while they assign sub-queries to different experts in a rules-based manner, our method processes the entire query and applies dynamic routing.

\section{Router Retriever}

\begin{algorithm}
\caption{Constructing Pilot Embedding Library}
\label{alg:pilot_embedding_library} 
\begin{algorithmic}[1]
\REQUIRE Domain-specific training datasets \( D_1, \dots, D_T \), experts \(\mathcal{E} = \{e_1, \dots, e_T\}\)
\STATE Initialize empty map \(\mathcal{P} = \{\}\) for the pilot embedding library

\FOR{each dataset \( D_i \) in \( \{D_1, \dots, D_T\} \)}
    \STATE Initialize an empty list \(\mathcal{L}_i = [\ ]\)
    \FOR{each instance \( x_j \) in \( D_i \)}
        \STATE \( e_{\text{max}} = \arg\max_{e_i \in \mathcal{E}} \text{Performance}(e_i, x_j) \) 
        \STATE Add pair \( (x_j, e_{\text{max}}) \) to \(\mathcal{L}_i\)
    \ENDFOR
    
    \FOR{each expert \( e_m \) in \(\mathcal{E}\)}
        \STATE \({Group}_m = \{x_j \mid e_{\text{max}} = e_m \text{ for } (x_j, e_{\text{max}}) \text{ in } \mathcal{L}_i\}\)
        
        \IF{\({Group}_m \) is not empty}
            \STATE \( \mathbf{c}_m = \text{Centroid}( \text{BaseEncoder}({Group}_m) \))  
            \STATE \(\mathcal{P}[e_m]\).append(\(\mathbf{c}_m\))
        \ENDIF
    \ENDFOR
\ENDFOR

\STATE \textbf{Output:} Pilot embeddings library \(\mathcal{P}\)
\end{algorithmic}
\end{algorithm}

In this section, we introduce \ours, a retrieval model composed of a base retrieval model and multiple domain-specific experts. As shown in Figure~\ref{fig:ours}, for a given input query, \textcircled{1} the most appropriate embedding is selected using a routing mechanism. Then, \textcircled{2} the query embedding is generated by passing the query through the selected expert alongside the base encoder. 

During training, we fix the base retrieval model and only finetune the specialized experts, one for each target domain using domain-specific training data. 
We use Contriever~\citep{izacard2021unsupervised} as our base encoder, and our experts are parameter-efficient LoRA~\citep{Hu2021LoRALA} modules.
We also pre-compute for each domain a set of representative \emph{pilot embeddings} that will help us route queries to appropriate experts.
We refer to the mappings between pilot embeddings to the associated trained expert for each domain as the \emph{pilot embedding library}.
This overall process is only performed once.
During inference, when given an input query, a \emph{routing mechanism} determines the appropriate expert by calculating the similarity score between the input query embedding and the pilot embeddings in the pilot embedding library, and then choosing the expert with the highest average similarity score.
This design allows for the flexible addition or removal of domain-specific experts without requiring any further training of the routing mechanism. 
To get into specifics:

\paragraph{Experts}
For each domain \( D_{i} \), where \( i = 1, \ldots, T \) and \( T \) is the total number of domains, we train an expert LoRA module \( e_{i} \) using the corresponding domain dataset. After the training step, we have a total of \( T \) different experts, \( \mathcal{E} = \{ e_1, e_2, \ldots, e_T \} \), with each expert \( e_i \) specialized for a specific domain.

\paragraph{Pilot Embedding Library}

To construct the pilot embedding library, given a domain-specific training dataset \( D_{i} = \{x_1, \dots, x_n\} \) where $x_j$ is an instance in $D_i$, we perform inference using all experts \(\mathcal{E}\) to identify which expert provides the most suitable representative embedding for each instance as shown in Algorithm~\ref{alg:pilot_embedding_library}. For each instance \( x_j \), we select \( e_{\text{max}} \), the expert that demonstrates the highest performance, defined as \( e_{\text{max}} = \arg\max_{e_i \in \mathcal{E}} \text{Performance}(e_i, x_j) \). This process produces pairs \((x_j, e_{\text{max}})\) for all instances in the dataset \( D_i \).
The intuition here is that $e_{max}$ for $x_j$ doesn't have to be the expert trained on the source $D_i$ that contains $x_j$. 

Next, with the constructed pairs \((x_j, e_{\text{max}})\), we group them by the ones that have the same \( e_{\text{max}} \).
This results in \( T \) groups, one for each domain (${Group}_m, m=1,\cdots, T$), where each ${Group}_m$ contains list of instances $x_j$ all sharing the same $e_{max}$.
If the ${Group}_m$ is not empty, we extract all embeddings for instances in the group using the base encoder (BaseModel), and
calculate the average, or centroid, embedding \(\mathbf{c}_m\), which is taken as the pilot embedding for the domain\footnote{We also experiment with $k$-means clustering of different numbers of $k$, the number of centroid embeddings, and having a single centroid embedding ($k$=1) yields the highest performance, as additional centroids often act as distractors. Details are in Appendix~\ref{app: c1}}.
This results in one pilot embedding per group, yielding a maximum of \( T \) pilot embeddings for the training dataset \( D_i \). Each of these embeddings is associated with a different expert, representing the most suitable one for that domain. 
When ${Group}_m$ is empty, we skip this step, so the number of pilot embeddings for $D_i$ could be less than $T$.

By repeating this process across all domain-specific training datasets \( D_1, \dots, D_T \), we obtain a maximum of \( T^2 \) pilot embeddings: $T$ domain-specific training datasets times $T$ groups per dataset.

\paragraph{Routing Mechanism}

Given an input query, we calculate the similarity between the query embedding extracted from the base encoder and the \( T^2 \) pilot embeddings in the \text{pilot embedding library}. We then average the similarity scores for $T$ pilot embeddings associated with the same expert, resulting in a mean similarity score for each expert. The expert corresponding to the highest mean similarity score is selected as the most suitable embedding model.

\section{Experimental Setup}
\paragraph{Baselines}
We compare the performance of \ours with a single base encoder model trained on the same dataset in a Multi-Task manner and a single base encoder model trained on a large-scale general-domain dataset MSMARCO. Following previous works~\citep{muqeeth2024learning, jang2023exploring}, we also evaluate two oracle settings: DatasetOracle and InstanceOracle\footnote{DatasetOracle and InstanceOracle correspond to Best Individual and Oracle, respectively, in prior works \citet{jang2023exploring} and \citet{muqeeth2024learning}}. The DatasetOracle setting is a dataset-level oracle that routes all queries in a dataset to the expert with the highest average performance for that dataset, while the InstanceOracle setting is an instance-level oracle that routes each individual instance to its best-performing expert. 

We also conduct experiments with various other routing techniques commonly used in language modeling tasks: ExpertClassifierRouter~\citep{shen2024learning}, ClassificationHeadRouter~\citep{muqeeth2024learning}, and DatasetRouter~\citep{ye2022retrieval, jang2023exploring}. 
ExpertClassifierRouter employs a binary classifier for each expert to calculate the probability of selecting that expert. The expert with the highest probability is chosen for the final selection. 
ClassificationHeadRouter uses a single classifier layer to determine the appropriate expert for each instance. 
DatasetRouter is the most similar to \ours, as it selects the expert by retrieving the instance with the highest similarity score. However, there are two key differences: \ours uses the predicted label, whereas DatasetRouter relies on the original dataset label. Also, \ours incorporates a clustering step to group instances, while DatasetRouter randomly samples 100 instances from the training dataset.
Further details of baselines and training methods are in Appendix~\ref{app: a1}. %supplementary materials.

\begin{figure}[t!]
    \centering
\begin{minipage}[b]{\linewidth}
    \centering
    \begin{minipage}[t]{0.5\linewidth}
        \centering
        \includegraphics[width=\linewidth]{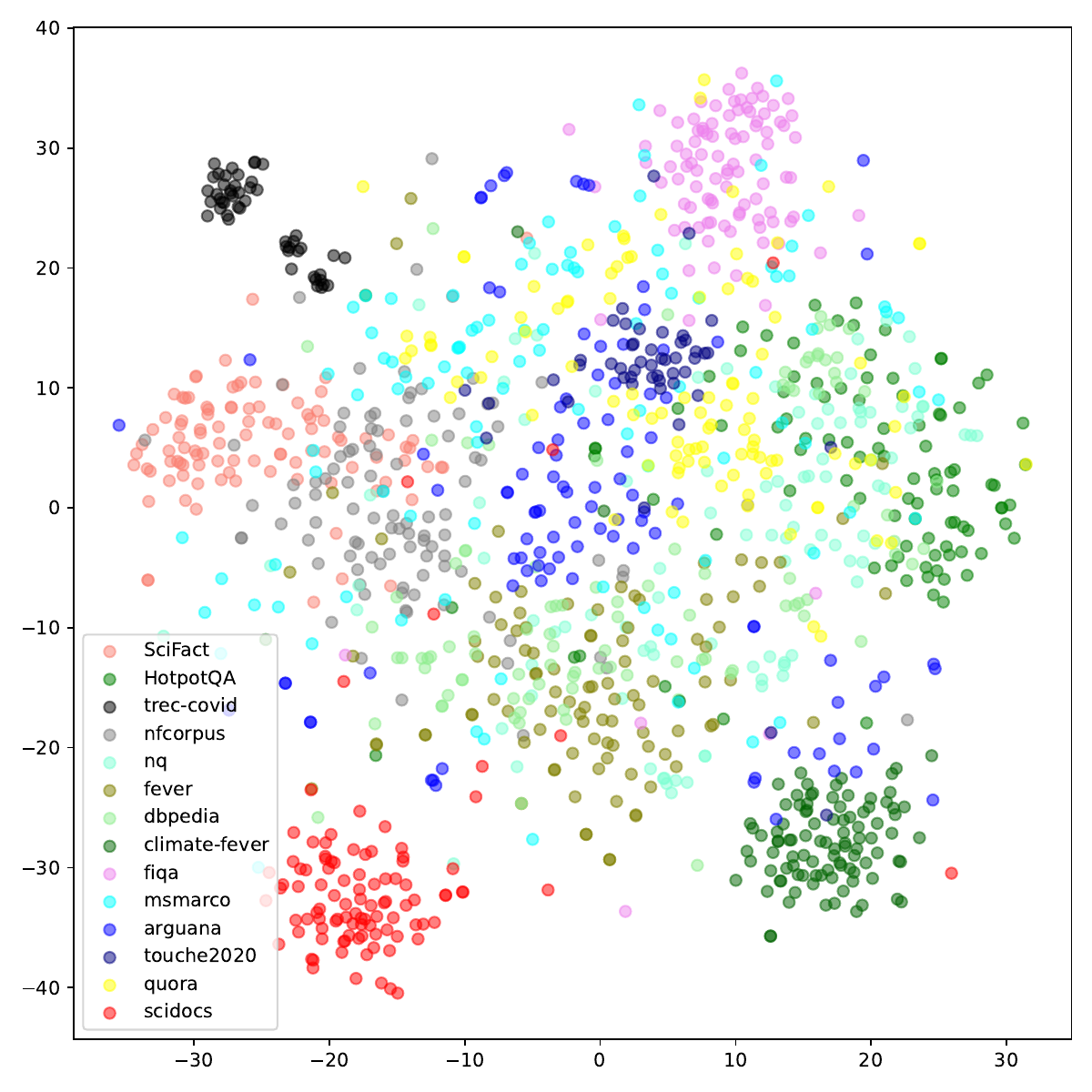}
    \end{minipage}\hfill
    \begin{minipage}[t]{0.5\linewidth}
        \centering
        \includegraphics[width=\linewidth]{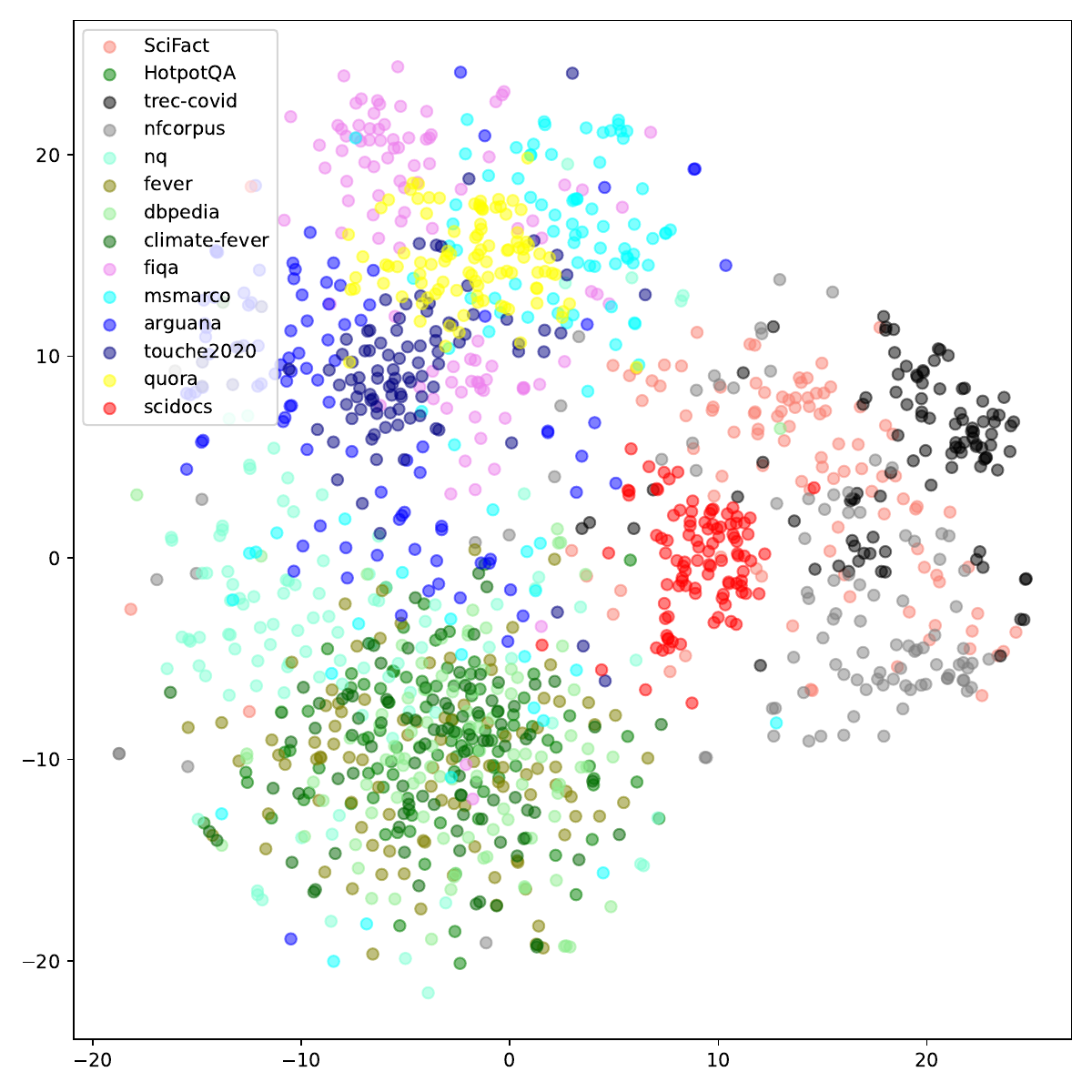}
    \end{minipage}
\end{minipage}
\caption{TSNE visualization of contriever embeddings for queries (left) and contexts (right) when sampled 100 instances from each dataset. 
We see high dispersion ``general-domain'' datasets like ArguAna and MSMARCO (blue) while ``domain-specific'' datasets like HotPotQA (green), NFCorpus (grey), SciFact (pink), and FiQA (purple) are tightly clustered. Datasets like Quora (yellow) have disperse queries but compact contexts.}
\label{fig:contriever_embeddings}
\end{figure}

\paragraph{Dataset}
We use the provided training\footnote{
In our early experiments, we noted that some training datasets in BEIR were so small that domain-specific models were underperforming MSMARCO on those target domains simply due to a lack of training data. To conduct a proper study of routing over experts, we had to first ensure that the respective experts were reasonably well-trained. 
As such, for our experiments, we also use the generated queries provided by BEIR at \url{https://huggingface.co/BeIR}.
} and test sets in the BEIR benchmark~\cite{thakur2021beir}.
We inspect BEIR domain splits using embeddings from our base encoder, a pre-trained Contriever model, in Figure~\ref{fig:contriever_embeddings}. 
We observe that datasets like MSMARCO~\cite{Campos2016MSMA} and ArguAna~\cite{Wachsmuth2018RetrievalOT} tend to have widely dispersed embeddings, indicative of their ``general-domain'' nature, while other datasets like HotpotQA~\cite{Yang2018HotpotQAAD}, NFCorpus~\cite{Boteva2016AFL}, SciFact~\cite{Wadden2020FactOF}, and FiQA~\cite{Maia2018WWW18OC} tend to have compact and tightly clustered instances, indicative of their ``domain-specific'' nature.
Some like Quora~\cite{quora_question_pairs_2017} have high dispersion in their queries but have tightly clustered contexts.
Where needed, we may use acronyms for datasets: ArguAna (AR), Quora (QU), MSMARCO (MS),  HotpotQA (HO), SciFact (SF), NFCorpus (NF), FiQA (FI), SciDocs (SD), and TREC-COVID (TR).

\begin{table*}[t!]
\centering
\fontsize{7}{10} \selectfont
    \begin{tabular}{ll|c|c|c|c|c|c|c|c}
    \toprule
     & & \text{MSMARCO} & \text{Quora} & \text{ArguAna} & \text{HotpotQA} & \text{NFCorpus} & \text{SciFact} & \text{FiQA} & Avg \\
    \midrule
    \multirow{3}{*}{\begin{tabular}[c]{@{}l@{}}Retrievers 
    \end{tabular}}
    & Single model on MSMARCO & \textbf{25.7} & \textbf{84.1} & 37.2 & 57.6 & 31.7 & 67.2 & 28.8 & 47.5 \\ 
    & Single model with Multi-Task & 22.4 & 82.0 & 36.9 & 52.1 & 32.9 & 69.4 & 28.9 & 46.4 \\
    & \ours (w/o MSMARCO expert) & 22.2 & 83.6 & \textbf{39.5} & \textbf{59.5} & \textbf{33.4} & \textbf{76.0} & \textbf{30.5} & \textbf{49.3} \\ 
    \midrule
    \multirow{4}{*}{Routing}
    & ExpertClassifierRouter & \textbf{23.8} & 82.5 & 37.9 & 53.1 & 31.5 & 67.1 & 29.1 & 46.4\\
    & ClassificationHeadRouter & 22.6 & 83.4 & 38.5 & 52.8 & 32.7 & 69.6 & 28.2 & 46.8 \\
    & DatasetRouter & 23.6 & \textbf{83.9} & 37.3 & 58.4 & 33.1 & 73.4 & 29.9 & 48.5 \\
    & \ours (w/ MSMARCO expert) & 23.0 & 83.8 & \textbf{38.6} & \textbf{59.9} & \textbf{33.4} & \textbf{77.6} & \textbf{30.8} & \textbf{49.6} \\
    \midrule
    \multirow{2}{*}{Oracles} & DatasetOracle & 25.7 & 84.5 & 40.2 & 59.9 & 34.4 & 79.8 & 32.2 & 50.9 \\
    & InstanceOracle & 34.5 & 89.9 & 48.5 & 66.6 & 39.0 & 85.4 & 39.6 & 57.6 \\
    \bottomrule
    \end{tabular}
\caption
     {\textbf{Retrievers:} When trained on the same dataset size, \ours consistently outperforms single model baselines (MSMARCO and Multi-Task) in terms of nDCG@10 on BEIR benchmark.
     \textbf{Routing:} \ours also surpasses various standard routing techniques commonly used in language modeling.
     \textbf{Oracles:} \ours achieves performance comparable to the DatasetOracle model. InstanceOracle indicates room for future work in router improvements.
     } 
\label{table: router_retriever.id}
\end{table*}

\paragraph{Hyperparameters}

We use the pre-trained Contriever~\citep{izacard2021unsupervised} as our base encoder and train experts~(LoRA) according to the settings in \citet{lee2023back}, with a rank of 8, an alpha of 32 per expert, thereby training approximately 0.5\% of the parameters (about 1M parameters) per expert. 
For training, we adopt the few-shot hyperparameters from \citet{izacard2021unsupervised}: a learning rate of 1e-4, a batch size of 256 with in-batch negatives, and a maximum of 500 epochs with early stopping. 
For brevity, we focus on presenting results for which experts are applied only to the query encoder, keeping the context encoder frozen. 
We include the results of applying experts to the context encoder in Appendix~\ref{app: b4}. %the supplementary materials.

\section{Results}

\subsection{Overall Performance}

Table~\ref{table: router_retriever.id} shows the performance of \ours using seven domain-specific experts compared to baseline models, evaluated on test sets of corresponding experts.
\textbf{\ours outperforms both single model baselines}---Multi-Task training over the same training data mix as well as training only on MSMARCO---even without an MSMARCO expert.\footnote{
For fair comparison, we ensure the number of training instances of used \ours and Multi-Task mix does not exceeds the number of training instances in MSMARCO.
}
And including an MSMARCO expert further improves performance for most domains.
These results underscore the importance of having separate embedding models (experts) for each domain and dynamically selecting the most appropriate expert for each query rather than relying on a single model to handle multiple domains. 
We include additional results with different combinations of experts in Appendix~\ref{app: b5}.

\subsection{Comparing Different Routing Techniques}

We experiment with different routing techniques commonly used in language modeling and compare them to our proposed routing mechanism. Results in Table~\ref{table: router_retriever.id} show that \textbf{the routing technique used in \ours consistently achieves the highest performance}. 
In fact, ClassificationHeadRouter and ExpertClassifierRouter approaches tend to underperform compared to simply using a single retriever trained solely on MSMARCO. DatasetRouter, which is the closest to \ours, tends to show higher performance than the single model trained on MSMARCO but also often still shows lower performance than \ours. 
These results suggest that \textbf{routing techniques developed for language modeling may not generalize well to information retrieval}. We hypothesize that the differences in the effectiveness of routing techniques between language modeling and information retrieval can be explained from the following perspective: 
In language modeling, routing decisions are often made at the token level, which allows for greater flexibility and reduces the impact of any single choice. However, in information retrieval, where a single representative embedding is required, the choice of expert is made only once per instance. This makes the process more vulnerable to the routing technique used and thus requires greater care to ensure precise routing.
We achieve this by designing a routing mechanism around embedding similarities, which leans into the strengths of our encoder models.
We are excited to see more sophisticated routing methods inspired by retrieval-specific designs in future works, especially closing the gap to InstanceOracle performance.

\begin{table}[t!]
\centering
\fontsize{7}{10} \selectfont
    \begin{tabular}{c|cc}
    \toprule
    & w/ Experts & w/o Experts \\
    \midrule
    Single model on MSMARCO & 47.5 & 31.6 \\
    Single model with Multi-Task & 46.4 & 31.2 \\
    \ours (w/ MSMARCO expert) & \textbf{49.6} & \textbf{31.9} \\
    \midrule 
    DatasetOracle & 50.9 & 34.2 \\
    InstanceOracle & 57.6 & 41.5 \\
    \bottomrule
    \end{tabular}
\caption
     {\ours not only shows high performance (nDCG@10) for datasets with trained experts but it also generalizes to those without experts. ``w/ Experts'' results are taken from Table~\ref{table: router_retriever.id}. ``w/o Experts'' averages results on seven other BEIR test sets that lack training sets.
     } 
\label{table: general_perf}
\end{table}

\subsection{Zero-shot Generalization to Unseen Domains}

We've demonstrated \ours's effective use of domain-specific training data, but what of unseen test sets that don't have corresponding trained experts? 
In Table~\ref{table: general_perf}, we evaluate our models for true zero-shot generalization to seven more BEIR test sets: Touche-2020~\cite{touche}, Climate-FEVER~\cite{Diggelmann2020CLIMATEFEVERAD}, DBPedia~\cite{Hasibi2017DBpediaEntityVA}, NaturalQuestions~\cite{Kwiatkowski2019NaturalQA}, FEVER~\cite{Thorne2018FEVERAL}, TREC-COVID~\cite{Roberts2020TRECCOVIDRA}, and SciDocs~\cite{Cohan2020SPECTERDR}.
We find \textbf{the benefits of \ours extend beyond the datasets for which specific experts were trained}. We provide dataset-specific results in Appendix~\ref{app: b7}.

\subsection{Training and Inference Efficiency}
\ours achieves high training efficiency by using parameter-efficient LoRA experts, which account for only 0.5\% of the parameters per expert. 
This makes the addition of new experts insignificant in terms of total parameter count.
It uses the same amount of training data as any multi-task approach.
However, unlike multi-task training which requires retraining the entire model when adding, removing, or changing domains, \textbf{\ours allows for these modifications without additional training}, as our routing technique is training-free.
However, during inference, computing the query embedding involves two forward passes: first to identify the appropriate expert (routing), and second to generate the final query embedding. Improving the computation efficiency of this routing technique is a direction for future work. Detailed analysis over the efficiency in Appendix~\ref{app: b8}.

% ---------------------- ANALYSIS --------------

\section{Analysis}

\subsection{Impact of Dataset Size when Training Experts}

\begin{figure}[t!]
    \centering
\begin{minipage}[b]{\linewidth}
    \centering
    \begin{minipage}[t]{\linewidth}
        \centering
        \includegraphics[width=\linewidth]{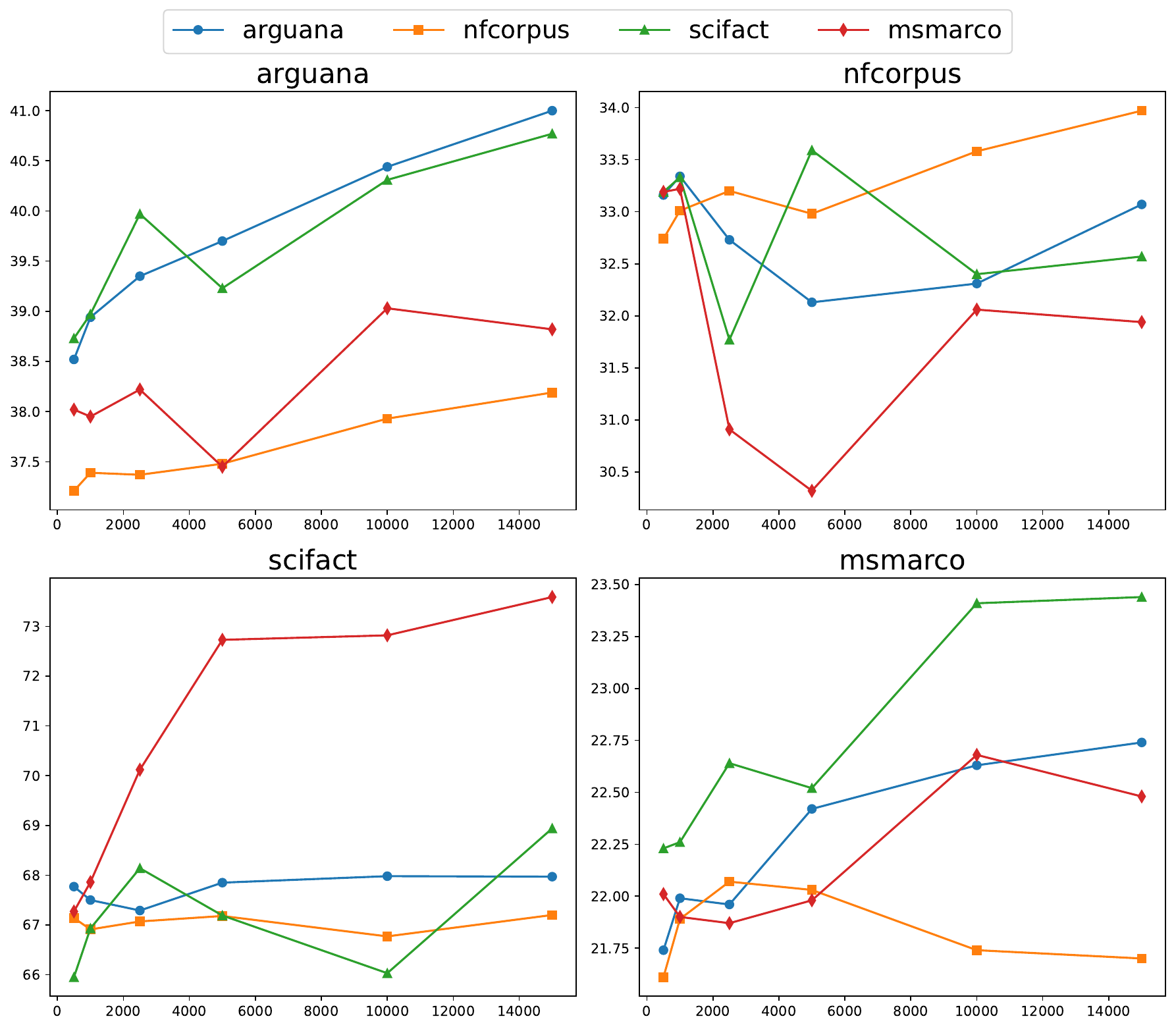}
    \end{minipage}
\end{minipage}
\caption{Single expert performance (nDCG@10; y-axis) against number of training instances (x-axis). Each line color represents the training dataset used, and each plot is a BEIR test dataset.
As we increase training set size, in-domain performance increases rapidly, but may not transfer to improved out-of-domain performance.
} 
\label{fig:dataset2performance}
\end{figure}

Figure~\ref{fig:dataset2performance} shows the relationship between the amount of training data and the performance of a single expert retriever.
For in-domain evaluation datasets, performance generally improves as the number of training instances increases. However, in out-of-domain evaluation datasets, simply increasing the number of training samples does not necessarily lead to better performance.
Interestingly, when testing out-of-domain, experts perform better when trained on general domains (e.g., ArguAna and MSMARCO) compared to domain-specific experts (e.g., SciFact and NFcorpus). We attribute this to the broader coverage of general-domain datasets, as illustrated in Figure~\ref{fig:contriever_embeddings}. 
These results suggest that while a larger training dataset is generally beneficial for expert in-domain performance, broad coverage and diversity of the training dataset have a more significant impact on out-of-domain performance. More details on the performance of each expert are in Appendix~\ref{app: b1}.

\begin{figure}[t!]
    \centering
\begin{minipage}[b]{\linewidth}
    \centering
    \begin{minipage}[t]{0.85\linewidth}
        \centering
        \includegraphics[width=\linewidth]{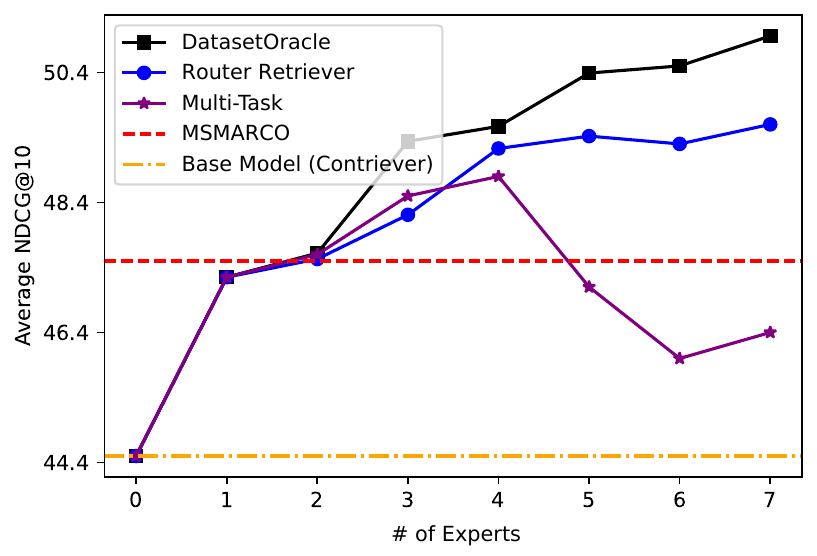}
    \end{minipage}
\end{minipage}
\caption{Average nDCG@10 (y-axis) by the number of experts (x-axis) for various models. \ours tends to show improved performance as the number of experts increases, outperforming a single MSMARCO-trained model even with just three experts despite less training data.} 
\label{fig:perf_by_gatenum}
\end{figure}

\begin{figure}[t!]
    \centering
\begin{minipage}[b]{\linewidth}
    \centering
    \begin{minipage}[t]{0.85\linewidth}
        \centering
        \includegraphics[width=\linewidth]{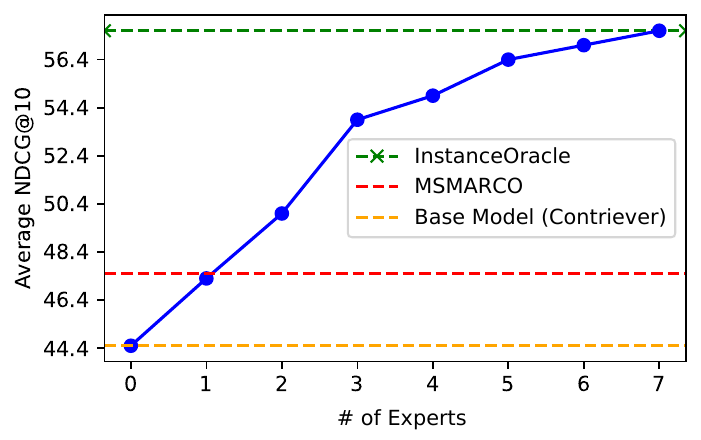}
    \end{minipage}
\end{minipage}
\caption{Average instance-level oracle routing performance nDCG@10 (y-axis) by the number of available experts (x-axis). The improvement rate tends to be high when adding experts initially followed by diminishing returns.} 
\label{fig:oracle_perf_by_gates}
\end{figure}

\subsection{Impact of Number of Experts}
Figure~\ref{fig:perf_by_gatenum} shows the relationship between the number of experts and the performance of \ours.
It outperforms the single MSMARCO-trained model even with just three experts, indicating that despite not having as diverse or large a training dataset as MSMARCO, the advantage of having multiple embedding models and the ability to select the most suitable one leads to better performance. 
The performance of multi-task training tends to fluctuate as the number of domains (experts) increases. We hypothesize that with a large number of domains, the model struggles to find the optimal embedding for general cases due to high variance across training datasets.

Yet, Figure~\ref{fig:perf_by_gatenum} also shows diminishing returns in \ours as we increase the number of experts, but a consistent increase in DatasetOracle.
To further study whether this is due to the need for better routing, we experiment with InstanceOracle but varying the pool of available experts. 
Figure~\ref{fig:oracle_perf_by_gates} shows as the number of experts increases, InstanceOracle performance also improves quickly before encountering some diminishing returns. 
In repeating these experiments in Appendix~\ref{app: c2}, we found this is true regardless of the expert combinations or order in which they're added.
Overall, we interpret the results to mean that \ours's routing technique tends to be more distracted as more experts are added, which could motivate future work on scaling router fidelity closer to InstanceOracle to handle the higher complexity of more experts.

Table~\ref{table: sequential_perf} shows results from experiments in which we sequentially add experts to \ours.
At low expert counts, each addition of a new expert can dramatically change performance across tasks that previously had an in-domain expert.
For example, adding a HotpotQA expert caused performance on ArguAna to drop from 40.1 to 38.5 and SciFact from 76.7 to 72.2, while causing the expected improvement in HotpotQA to increase from 55.3 to 59.2.
At higher expert counts, these side-effects are much more muted.
For example, adding SciDocs or TREC-COVID experts to a seven-expert \ours improves performance for SciDocs (14.8 to 16.3) and TREC-COVID (44.9 to 56.2), but doesn't change overall performance for the other categories.

\begin{figure}[t!]
\centering
    \begin{subfigure}[t]{\linewidth}
        \centering
        \includegraphics[width=0.9\linewidth]{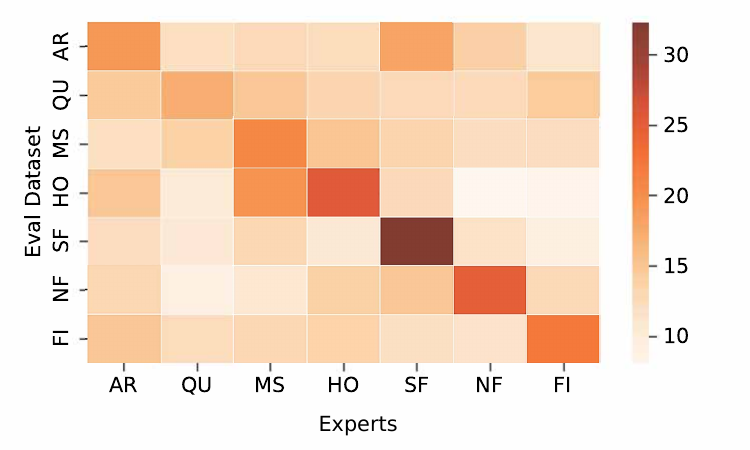}
        \caption{InstanceOracle}
        \label{fig:max_gate_rate}
    \end{subfigure}
    
    \begin{subfigure}[t]{\linewidth}
        \centering
        \includegraphics[width=0.9\linewidth]{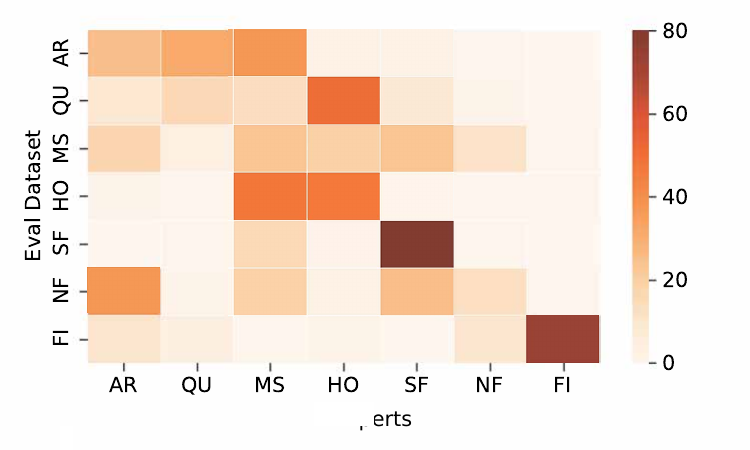}
        \caption{\ours}
        \label{fig:routing_result}
    \end{subfigure}
    
\caption{
For each evaluation dataset (y-axis), how often is each expert chosen (x-axis)? Darker cells mean more frequent selection. Diagonal entries mean in-domain selection.
\textbf{(a)} While ``general'' experts like AR and MS appear well-suited to tackle instances from other datasets (darker columns), instances from certain datasets like SF and NF must be routed to the in-domain expert (sparse columns with single dark concentration). 
\textbf{(b)} \ours has sparser routing behavior, tending towards following dataset boundaries, which explains similar results as DatasetOracle.}
\label{fig:combined_routing}
\end{figure}

\begin{table*}[t!]
\centering
\fontsize{7}{10} \selectfont
    \begin{tabular}{l|l|ccccccccc}
    \toprule
    \textbf{Start} & \textbf{Addition} & \textbf{AR} & \textbf{NF} & \textbf{SF} & \textbf{FI} & \textbf{HO} & \textbf{QU} & \textbf{MS} & \textbf{SD} & \textbf{TR}\\
    \midrule
    \multicolumn{9}{l}{Unstable additions with few experts} \\
    \midrule
    AR, NF, SF, FI & - & 40.1 & 32.3 & 76.7 & 30.7 & 55.3 & 83.2 & 22.1 & 15.1 & 43.1 \\
    AR, NF, SF, FI & + HotpotQA & 38.5 & 33.0 & 72.2 & 27.9 & 59.2 & 82.7 & 22.3 & 15.9 & 43.6 \\
    AR, NF, SF, FI, HO & + Quora &  39.5 & 33.4 & 76.0 & 30.5 & 59.5 & 83.6 & 22.2 & 15.1 & 44.3 \\
    \midrule 
    \multicolumn{9}{l}{Stable additions with more experts} \\
    \midrule
    AR, NF, SF, FI, HO, QU & + MSMARCO & 38.6 & 33.4 & 77.6 & 30.8 & 59.9 & 83.8 & 23.0 & 14.8 & 44.9 \\
    AR, NF, SF, FI, HO, QU, MS & + SciDocs & 38.8 & 32.7 & 76.9 & 30.3 & 59.8 & 84.1 & 22.9 & 16.3 & 44.7 \\
    AR, NF, SF, FI, HO, QU, MS & + TREC-COVID & 38.4 & 32.7 & 77.3 & 31.4 & 59.9 & 83.9 & 22.7 & 14.6 & 56.2 \\
    \bottomrule
\end{tabular}
\caption
     {
     Performance (nDCG\@10) across BEIR datasets while sequentially adding more experts to \ours.
     } 
\label{table: sequential_perf}
\end{table*}

\subsection{Analyzing Routing Errors using InstanceOracle}
In Figure~\ref{fig:combined_routing}, we hope to understand the gap in performance between \ours routing and InstanceOracle routing by inspecting which instances are routed to which experts by either method.
First, Figure~\ref{fig:max_gate_rate} shows that general-domain datasets like ArguAna and MSMARCO yield experts that can handily tackle instances from other diverse datasets; hence, they receive an even distribution of queries from InstanceOracle.
However, for domain-specific datasets like SciFact or NFCorpus, the best performance is typically achieved by the expert trained specifically on that domain. 
Figure~\ref{fig:routing_result} then shows that \ours's expert selection tends to be much sparser, prioritizing routing instances to their source dataset's expert. 
This explains the similar performance seen in Table~\ref{table: router_retriever.id} between \ours and DatasetOracle, and motivates future work on more powerful routing mechanisms.
We add a detailed error analysis of our routing technique in Appendix~\ref{app: b6}.

\section{Conclusion}

In this paper, we introduce \ours, a retrieval model that leverages a mixture of domain-specific expert embeddings, guided by a routing mechanism to select the most suitable embedding for each query.
This approach is both lightweight and flexible, allowing for the addition or removal of experts without additional training. Our experiments demonstrate that it consistently outperforms single embedding models, showcasing the advantages of integrating domain-specific experts. Additionally, it surpasses various widely used routing techniques in language modeling, emphasizing the significance of effective routing for information retrieval tasks. Our results highlight the crucial role of domain-specific experts in improving retrieval performance across diverse domains. 
Yet, there remains much more room for improvement, as indicated by the higher performance of an instance-level oracle router compared to our method.
We hope our work spurs the broader research community to search for more powerful methods for training expert retrievers and combining them with efficient routing techniques.

\section*{Acknowledgments}
We thank Nandan Thakur, Orion Weller, Jiyeon Kim, Hanseok Oh, and the Semantic Scholar Research team at Ai2 for helpful discussions and constructive feedback.

\bibliography{aaai25}

\appendix

\newpage
\section{Experimental Setup} \label{app: a}
\subsection{Baselines} \label{app: a1}
\paragraph{MSMARCO}
This baseline uses a single MSMARCO gate, which is trained on a large-scale, general-domain dataset without any routing techniques applied.
\paragraph{Multi-Task}
In this approach, we train a single embedding model on all datasets simultaneously in a multi-task manner. We keep the number of training datasets for each label the same, keeping to the one with the minimum value by sampling. 
\paragraph{Best Individual}
This represents the oracle performance when selecting the single best-performing gate for \textit{each dataset}. For example, if the SciFact gate shows the highest overall performance on the SciDocs evaluation dataset compared to other gates, the performance of the SciFact gate is recorded as the best individual performance for SciDocs.
\paragraph{Oracle}
This is the oracle performance when selecting the best-performing gate for \textit{each individual instance}. For example, within the SciDocs dataset, certain instances might achieve the highest performance with the SciFact gate, while others might perform better with the MSMARCO gate. This baseline measures the performance when, for each instance, the gate that yields the best result is selected.
\paragraph{ExpertClassifierRouter}
This routing technique, inspired from \citet{shen2024learning}, uses a binary classifier for each gate. For each instance, the classifier calculates the probability of selecting or not selecting a specific gate. The gate with the highest probability of being selected is chosen.

To construct the training dataset, we use the predicted label ($g_\text{max}$) from the \textbf{Pilot Embedding Library}. For each ($x_i$, $g_\text{max}$) pair, we randomly sample instances where the maximum gate differs, which are used to train the "not choosing the gate" label. The dataset is balanced across labels, with the following number of training instances for each dataset: AR (16,108), FI (1070), SF (1,414), NF (892), HO (4,618), QU (4,326), and MS (4,252). Please note that the training datasets only consist of instances where only a single gate shows maximum performance.
We then train a binary classifier for each gate to predict whether an instance is likely to achieve the highest performance through that gate.

\paragraph{ClassificationHeadRouter}
This routing technique, inspired from \citet{muqeeth2024learning}, uses a classification head where the number of labels corresponds to the number of gates. The gate with the highest predicted probability is selected as the one likely to yield the best performance. To ensure balance, we equalize the number of training instances for each label, matching the dataset with the fewest instances (NFcorpus with 892 instances, other numbers in ExpertClassifierRouter paragraph). AS a result, the total number of training instances is 6,244.

\paragraph{DatasetRouter}
This routing technique, inspired from \citet{ye2022retrieval, jang2023exploring}, is the closest baseline to \ours. It samples 100 training instances from each dataset and when given a query, it retrieves the most relevant instances from these samples. The gate trained on the dataset from which the sample originated is then used. 

The key differences between DatasetRouter and \ours are as follows. (1) \ours uses the predicted label to map an instance to a gate, while DatasetRouter relies on the original dataset label. For example, if a training instance from MSMARCO performs best with the sciFact gate, \ours will select the Scifact gate for a similar query, whereas DatasetRouter will select the MSMARCO gate. (2) \ours incorporates a clustering step, grouping similar instances together and using centroid embeddings, rather than treating each instance individually.

\subsection{Datasets} \label{app: a2}
\begin{table*}[t!]
\centering
\fontsize{6.5}{10}\selectfont
    \begin{tabular}{c|cccccc}
    \toprule
    Domain & Name & Task & Train~(k) & Gen Train~(k) & Test & Corpus~(k) \\
    \midrule
    \multirow{3}{*}{Misc.} 
    & ArguAna (AR)~\citep{Wachsmuth2018RetrievalOT} & Argument Retrieval & - & 23 &1,406 & 8.7 \\
    & Touche-2020 (TO)~\citep{Bondarenko2020OverviewOT} & Argument Retrieval & - & - & 49 & 382.5 \\
    & MSMARCO (MS)~\citep{Campos2016MSMA} & Passage-Retrieval & 503 & - &6,980 & 8,842 \\
    \midrule
     \multirow{4}{*}{Wikipedia} 
    & NaturalQuestions (NQ)~\citep{Kwiatkowski2019NaturalQA} & Question Answering & - & -& 3,452 & 2,681\\
    & HotpotQA (HO)~\citep{Yang2018HotpotQAAD} & Question Answering & 85 & - & 7,405 & 5,233 \\
    & DBpedia (DB)~\citep{Hasibi2017DBpediaEntityVA} & Entity-Retrieval &  - & - &400 & 4,636 \\    
    & FEVER (FE)~\citep{Thorne2018FEVERAL} & Fact Checking  & 110 & - &6,666 & 5,417 \\
    & Climate-FEVER (CL)~\citep{Diggelmann2020CLIMATEFEVERAD} & Fact Checking &- & -& 1,535 & 5,417 \\
    \midrule
    \multirow{2}{*}{Bio-Medical} 
    & TREC-COVID (TR)~\citep{Roberts2020TRECCOVIDRA} & Bio-Medical Retrieval & - &432&50 & 171\\
    & NFCorpus (NF)~\citep{Boteva2016AFL} & Bio-Medical Retrieval &  2.6 & 10.8 & 323 & 3.6 \\
    \midrule 
     \multirow{2}{*}{Scientific} 
    & SCIDOCS (SD)~\citep{Cohan2020SPECTERDR}  & Citation-Prediction  & -& 67 & 1,000   & 25.7 \\
    & SciFact (SF)~\citep{Wadden2020FactOF} & Fact Checking  & 0.8 & 15.4& 300 & 5.2\\
    \midrule
     \multirow{1}{*}{Finance} 
    & FIQA-2018 (FI)~\citep{Maia2018WWW18OC} & Question Answering & 5.5 & 162 &648 & 57.6 \\
    \midrule 
    \multirow{1}{*}{Quora} 
    &Quora (QU) & Duplicate-Question Retrieval &- & 200  &  10,000 & 523 \\ 
    \bottomrule
    \end{tabular}
\caption
     {Data statics of 14 datasets in BEIR benchmark. Units of the numbers of training dataset (Train), generated training dataset (Gen Train), and corpus are in thousands. For most datasets, training datasets are not provided and for some datasets, we failed to download the generated training dataset.}
\label{table: beir}
\end{table*}

\paragraph{Stats of Training Dataset}

Table~\ref{table: beir} presents the statistics and details of the datasets in the BEIR benchmark, which we used for training and evaluation. We sampled datasets from Quora to ensure that the number of training instances for AR, HO, NF, SF, FI, and QU matches that of MS.

\begin{table*}[t!]
\centering
\fontsize{7.5}{9}\selectfont
\caption{\fontsize{7.5}{10}\footnotesize Examples where the dataset label differs from the predicted label based on the highest-performing gate.} 
\begin{tabular}{ m{10cm} m{2cm} m{2cm}} 
    \toprule
    \textbf{Question} & \textbf{Dataset} & \textbf{Max Gate} \\
    \midrule
        APOE4 expression in iPSC-derived neurons results in decreased tau phosphorylation. & SciFact & NFCorpus \\
        which mir regulates the autophagy of cells & SciFact & NFCorpus \\
        what kind of leader should i be as the chief executive & FiQA-2018 & ArguAna \\
        what is casual dining dining & FiQA-2018 & HotpotQA \\
        is it better to be a vegan or vegetarian? & ArguAna & SciFact \\
        could we ban animal testing & ArguAna & SciFact \\
        why do humans eat meat & ArguAna & Quora \\
    \bottomrule
\end{tabular}
\label{table: oracle_ex}
\end{table*}

\paragraph{Examples of Oracle}
Table~\ref{table: oracle_ex} shows examples of questions where a gate from a different dataset outperforms the gate trained on the dataset to which the question belongs. We observe that questions related to biology often achieve higher performance with the NFCorpus gate, while those involving scientific knowledge tend to favor the SciFact gate, and questions requiring arguments perform better with the ArguAna gate. This pattern suggests that, even within a single dataset, some instances may be more closely aligned with other datasets, likely because the datasets were not labeled or constructed to avoid overlap with existing datasets.

\subsection{Hyperparameters} \label{app: a3}

We trained the Contriever model~\citep{izacard2021unsupervised} using an asymmetric architecture, where the query encoder encodes the query and the context encoder encodes the context. In our experiments, we fine-tuned only the LoRA (Low-Rank Adaptation) parameters of the query encoder, training approximately 1 million parameters per gate (which accounts for 0.5\% of the total model parameters).
For evaluation, we used the NDCG@10 metric, consistent with previous works~\citep{thakur2021beir, lee2023back}, which measures the ranking quality of the top 10 retrieved documents. All results were calculated using the official BEIR evaluation code.
The experiments were conducted on 8 or fewer A6000 GPUs (each with 40GB of memory). We utilized checkpoints from all pretrained models available on Huggingface\footnote{\url{https://huggingface.co/facebook/contriever}}. The experiments were performed over various combinations of gates, with all random seeds set to 10.

\section{Results} \label{app: b}

\begin{table*}[t!]
\centering
\fontsize{6.5}{10} \selectfont
    \begin{tabular}{c|cc|c|c|c|c|c|c}
    \toprule
    & \multicolumn{2}{c}{Misc} & \multicolumn{1}{c}{Wiki} & \multicolumn{1}{c}{Bio} & \multicolumn{1}{c}{Science} & \multicolumn{1}{c}{Quora} &\multicolumn{1}{c}{Finance}  \\
    \midrule
     & \text{AR} & MS & \text{HO} & \text{NF} & \text{SF} & \text{QU} & \text{FI} & Avg \\
    \midrule
    MSMARCO & 39.3 & \textbf{25.3} & 57.9 & 32.2 & 66.5 & \textbf{84.3} & 28.6 & 47.7 \\ 
    Multi-Task & 38.2 & 21.9 & 49.8 & 32.4 & 65.1 & 83.3 & 26.1 & 45.3 \\
    \ours & \textbf{40.5} & 21.0 & \textbf{61.3} & \textbf{32.7} & \textbf{68.2} & 82.5 & \textbf{30.0} & \textbf{48.0} \\
    \midrule
    Best Individual & 41.2 & 25.3 & 60.9 & 32.2 & 70.3 & 86.1 & 32.2 & 49.8 \\
    Oracle & 48.2 & 33.2 & 68.4 & 39.0 & 76.7 & 90.0 & 38.6 & 56.3 \\
    \bottomrule
    \end{tabular}
\caption
     {Performance of \ours when context encoder is trainable.} 
\label{table: router_retriever.id.context_trainable}
\end{table*}

\subsection{Unfreezing context encoder} \label{app: b4}

In our main experiments, we focus on scenarios where the context encoder is frozen, and only the LoRA of the query encoder is trainable to isolate the impact of routing on the query encoder alone. However, we observe that the overall performance trend remains similar even when the context encoder is not frozen, with the unfrozen models generally achieving higher performance. Table~\ref{table: router_retriever.id.context_trainable} presents the results when the context encoder is frozen. In these experiments, \ours consistently outperforms the MSMARCO-trained model and the Multi-Task model.

\begin{table}[t!]
\centering
\fontsize{6.5}{10} \selectfont
    \begin{tabular}{c|cc|c|c|c|c}
    \toprule
    & \multicolumn{2}{c}{Misc}  & \multicolumn{1}{c}{Bio} & \multicolumn{1}{c}{Science} & \multicolumn{1}{c}{Finance} & Avg \\
    \midrule
     & \text{AR} & MS & \text{NF} & \text{SF} & \text{FI} \\
    \midrule
    MSMARCO & 37.2 & 25.7 & 31.7 &  67.2  & 28.8 & 38.1  \\
    Multi-Task & 39.4 &21.2 &28.2 & 69.2 & \text{30.0} & 37.6 \\
    \ours & \textbf{40.1} & 22.1 & 32.3 & \textbf{76.7} & \textbf{30.7} & \textbf{40.4} \\
    \midrule
    \datawise & 40.2 & 22.4 & 34.4 & 79.8  & 32.2 & 41.8  \\
    \oracle & 47.5 & 29.3 & 38.0 & 84.5 & 37.4 & 47.3 \\
    \bottomrule
    \end{tabular}
\caption
     {\ours performance with four gates: AR, NF, SF, FI. Avg is an average performance over the dataset of all gates and MSMARCO.} 
\label{table: router_retriever.expert_4.id}
\end{table}

\begin{table}[t!]
\centering
\fontsize{6.5}{10} \selectfont
    \begin{tabular}{c|cc|c|c|c|c|c}
    \toprule
    & \multicolumn{2}{c}{Misc} & \multicolumn{1}{c}{Wiki} & \multicolumn{1}{c}{Bio} & \multicolumn{1}{c}{Science} &\multicolumn{1}{c}{Finance} & Avg \\
    \midrule
     & \text{AR} & MS & \text{HO} & \text{NF} & \text{SF}& \text{FI} \\
    \midrule
    MSMARCO & 37.2 & 25.7 &  57.6 & 31.7 & 67.2 & \textbf{28.8} & 41.4  \\
    Multi-Task & 37.7 & 22.0 & 58.6 &31.1 & 69.1& 28.4 & 41.2  \\
    \ours & \textbf{38.5} & 22.3 & \textbf{59.2} & \textbf{33.0} & \textbf{72.2} & 27.9 & \textbf{42.2} \\
    \midrule
    \datawise & 40.2 & 22.4 & 59.9 & 34.4 & 79.8 &  32.2 & 44.8 \\
    \oracle & 47.7 & 31.5 & 65.1 & 38.6 &84.8 & 38.4 & 51.0 \\
    \bottomrule
    \end{tabular}
\caption
     {\ours performance with five gates: AR, NF, SF, FI, HO. Avg is an average performance over the dataset of all gates and MSMARCO.} 
\label{table: router_retriever.expert_5.id}
\end{table}

\begin{table*}[t!]
\centering
\fontsize{6.5}{10} \selectfont
    \begin{tabular}{c|cc|c|c|c|c|c|c}
    \toprule
    & \multicolumn{2}{c}{Misc} & \multicolumn{1}{c}{Wiki} & \multicolumn{1}{c}{Bio} & \multicolumn{1}{c}{Science} & \multicolumn{1}{c}{Quora} &\multicolumn{1}{c}{Finance}  \\
    \midrule
     & \text{AR} & MS & \text{HO} & \text{NF} & \text{SF} & \text{QU} & \text{FI} & Avg \\
    \midrule
    MSMARCO & 37.2 &25.7 & 57.6 & 31.7 & 67.2 & \textbf{84.1} & 28.8 & 47.5  \\ 
    Multi-Task & 35.3 &21.5 & 55.3 & 32.3 & 65.3 & 82.8 & 29.3 & 46.0 \\
    \ours & \textbf{39.5} & 22.2 & \textbf{59.5} & \textbf{33.4} & \textbf{76.0} & 83.6 & \textbf{30.5} & \textbf{49.3} \\ 
     \midrule
    \datawise & 40.2 & 22.6 &59.9 & 34.4 & 79.8 & 84.5 & 32.2 & 50.5\\
    \oracle & 48.0 & 32.7 &65.5 & 38.8 & 85.0 & 89.7 & 39.2 & 57.0 \\
    \bottomrule
    \end{tabular}
\caption
     {\ours performance with six gates: AR, NF, SF, FI, HO, QU. Avg is an average performance over the dataset of all gates and MSMARCO.} 
\label{table: router_retriever.expert_6.id}
\end{table*}

\begin{table*}[t!]
\centering
\fontsize{6.5}{10} \selectfont
    \begin{tabular}{c|cc|c|c|c|c|c|c}
    \toprule
    & \multicolumn{2}{c}{Misc} & \multicolumn{1}{c}{Wiki} & \multicolumn{1}{c}{Bio} & \multicolumn{1}{c}{Science} & \multicolumn{1}{c}{Quora} &\multicolumn{1}{c}{Finance}  \\
    \midrule
     & \text{AR} & \text{MS} & \text{HO} & \text{NF} & \text{SF} & \text{QU} & \text{FI} & Avg \\
    \midrule
    MSMARCO & 37.2 & \textbf{25.7} & \textbf{57.6} & 31.7 & 67.2 & \textbf{84.1} & 28.8 &  47.5  \\
    Multi-Task & 36.9 & 22.4 & 52.1 & 32.9 & 69.4 & 82.0 & 28.9 & 46.4  \\
    \ours & \textbf{38.6} & 23.0 & \textbf{59.9} & \textbf{33.4} & \textbf{77.6} & 83.8 & \textbf{30.8} & \textbf{49.6} \\
    \midrule
    Best Individual & 40.2 & 25.6 & 59.9 & 34.4 & 79.8 & 84.5 & 32.2 & 50.9 \\
    Oracle &48.5 & 34.5 & 66.6 & 39.0 & 85.4 & 89.9 & 39.6 & 57.6 \\
    \bottomrule
    \end{tabular}
\caption
     {\ours performance with seven gates: AR, NF, SF, FI, HO, QU, MS} 
\label{table: router_retriever.expert_7.id}
\end{table*}

\subsection{Detailed numbers by gates} \label{app: b5}
In this section, we show detailed number of performance with different combinations of gates. 
Table~\ref{table: router_retriever.expert_4.id} shows performance with AR, NF, SF, FI as gates. 
Table~\ref{table: router_retriever.expert_5.id} shows performance with AR, \textbf{HO}, NF, SF, FI as gates.
Table~\ref{table: router_retriever.expert_6.id} shows performance with AR, HO, NF, SF, \textbf{QU}, FI as gates.
Table~\ref{table: router_retriever.expert_7.id} shows performance with AR, \textbf{MS}, HO, NF, SF, QU, FI as gates.
Figure 4 shows only with three gates, \ours outperforms the MSMARCO-trained ones thereby in all results, we can see that \ours outperforms the MSMARCO-trained ones and multi-task baselines.

\begin{table*}[t!]
\centering
\fontsize{6.5}{10} \selectfont
    \begin{tabular}{c|ccc|ccccc|cc|cc|c|c|c}
    \toprule
    & \multicolumn{3}{c}{Misc} & \multicolumn{5}{c}{Wiki} & \multicolumn{2}{c}{Bio} & \multicolumn{2}{c}{Science} & \multicolumn{1}{c}{Quora} &\multicolumn{1}{c}{Finance}  \\
    \midrule
     & \textbf{AR} & TO & MS & CL & DB & NQ & FE & HO & \textbf{NF} & TR & SD & \textbf{SF} & QU & \textbf{FI} & Avg \\
    \midrule
    MSMARCO & 37.2 & 18.3 & \text{25.7} & 16.0 & 32.7 & \text{29.3} & \text{68.8} & \text{57.6} & 31.7 & 41.2 & 14.6 & 67.2 & \text{84.1} & 28.8 & 39.5 \\
    Multi-Task & 39.4 & \text{18.9} & 21.2& \text{16.1}& 27.0& 23.5& 60.0&41.6 &28.2 &41.6 &15.0& 69.2& 82.2 & \text{30.0} &  36.7 \\ 
    \ours & \text{40.1} & 18.4 & 22.1 & 15.4 & \text{33.6} & 25.5 & 67.3 & 55.3 & \textbf{32.3} & \text{43.1} & \text{15.1} & \text{76.7} & 83.2 & \text{30.7} &  \textbf{39.9} \\
    \midrule
    \datawise & 40.2 & 18.5 & 22.4 & 15.7 & 31.0 & 26.2 & 68.9 &  54.2 & 34.4 & 44.6 & 15.5 & 79.8 & 83.7 & 32.2 & 40.5\\
    \oracle & 47.5 & 23.8 & 29.3 & 19.7 &36.5 & 33.5 & 76.7 & 59.4 & 38.0 & 52.5 & 20.0 & 84.5 & 88.1 & 37.4 & 46.2 \\
    \bottomrule
    \end{tabular}
\caption
     {\ours performance with four gates: AR, NF, SF, FI. Avg is an average performance over all datasets.} 
\label{table: router_retriever.expert_4.ood}
\end{table*}

\begin{table*}[t!]
\centering
\fontsize{6.5}{10} \selectfont
    \begin{tabular}{c|ccc|ccccc|cc|cc|c|c|c}
    \toprule
    & \multicolumn{3}{c}{Misc} & \multicolumn{5}{c}{Wiki} & \multicolumn{2}{c}{Bio} & \multicolumn{2}{c}{Science} & \multicolumn{1}{c}{Quora} &\multicolumn{1}{c}{Finance}  \\
    \midrule
     & \textbf{AR} & TO & MS & CL & DB & NQ & FE & \textbf{HO} & \textbf{NF} & TR & SD & \textbf{SF} & QU & \textbf{FI}& Avg \\
    \midrule
    MSMARCO & 37.2 & 18.3 & \text{25.7} & 16.0 & 32.7 & \text{29.3} & \text{68.8} & 57.6 & 31.7 & 41.2 & 14.6 & 67.2 & \text{84.1} & \text{28.8} & 39.5 \\
    Multi-Task & 37.7 &17.8 & 22.0& 15.4& 33.2& 26.6& 65.5& 58.6 &31.1 &41.6 &15.1 & 69.1& 82.7& 28.4  & 39.0 \\
    \ours & \text{38.5} & \text{19.8} & 22.3 & \text{17.1} & \text{35.4} & 27.8 & 67.4 & \text{59.2} & \text{33.0} & \text{43.6} & \text{15.9} & \text{72.2} & 82.7 & 27.9 &  \textbf{40.2} \\
    \midrule
    \datawise & 40.2 & 19.7 & 22.4 & 17.7 & 36.1 & 28.8 & 68.9 & 59.9 & 34.4 & 44.6 & 16.2 & 79.8 & 83.7 & 32.2 & 41.8\\
    \oracle & 47.7 & 25.7 & 31.5 & 21.3 & 40.4 &  37.9 & 79.5 & 65.1 & 38.6 & 53.1 & 20.7  &84.8  & 88.7 & 38.4  &48.1\\
    \bottomrule
    \end{tabular}
\caption
     {\ours performance with five gates: AR, NF, SF, FI, HO. Avg is an average performance over all datasets.} 
\label{table: router_retriever.expert_5}
\end{table*}

\begin{table*}[t!]
\centering
\fontsize{6.5}{10} \selectfont
    \begin{tabular}{c|ccc|ccccc|cc|cc|c|c|c}
    \toprule
    & \multicolumn{3}{c}{Misc} & \multicolumn{5}{c}{Wiki} & \multicolumn{2}{c}{Bio} & \multicolumn{2}{c}{Science} & \multicolumn{1}{c}{Quora} &\multicolumn{1}{c}{Finance}  \\
    \midrule
     & \textbf{AR} & TO & \textbf{MS} & CL & DB & NQ & FE & \textbf{HO} & \textbf{NF} & TR & \textbf{SD} & \textbf{SF} & \textbf{QU} & \textbf{FI} & Avg \\
    \midrule
    MSMARCO & 37.2 & \text{18.3} & \text{25.7} & \text{16.0} & 32.7 & \text{29.3} & \text{68.8} & 57.6 & 31.7 & 41.2 & 14.6 & 67.2 & \text{84.1} & \text{28.8} &  39.5 \\
    Multi-Task & 37.7 &17.8 & 22.0& 15.4& 33.2& 26.6& 65.5& 58.6 &31.1 &41.6 &15.1 & 69.1& 82.7& 28.4 & 39.0 \\
    \ours & \text{38.8} & 17.7 & 22.9 & 15.2 & 31.7 & 27.7 & 67.5 & 59.8 & 32.7 & 44.7 & 16.3 & 76.9 & 84.1 & 30.3 & \textbf{40.4} \\
    \midrule
    \datawise & 40.2 & 19.7 & 25.6 & 17.7 & 36.1 & 29.3 & 70.8 & 59.9 & 34.4 & 49.7 & 16.2 & 79.8 & 84.5 & 32.2 & 42.6 \\
    \oracle & 49.1 & 27.4 & 33.8 & 20.7 & 42.0 & 40.8 & 82.6 & 66.3 & 40.2 & 55.4 & 18.5 & 86.1 & 88.5 & 33.3 & 48.9\\
    \bottomrule
    \end{tabular}
\caption
     {\ours performance with eight gates: AR, NF, SF, FI, HO, MS, SD, QU. Avg is the average performance of all datasets.} 
\label{table: router_retriever.expert_8_sd}
\end{table*}

\begin{table*}[t!]
\centering
\fontsize{6.5}{10} \selectfont
    \begin{tabular}{c|ccc|ccccc|cc|cc|c|c|c}
    \toprule
    & \multicolumn{3}{c}{Misc} & \multicolumn{5}{c}{Wiki} & \multicolumn{2}{c}{Bio} & \multicolumn{2}{c}{Science} & \multicolumn{1}{c}{Quora} &\multicolumn{1}{c}{Finance} \\
    \midrule
     & \textbf{AR} & TO & \textbf{MS} & CL & DB & NQ & FE & \textbf{HO} & \textbf{NF} & \textbf{TR} & \text{SD} & \textbf{SF} & \textbf{QU} & \textbf{FI} & Avg \\
    \midrule
    MSMARCO & 37.2 & 18.3 & \text{25.7} & 16.0 & 32.7 & \text{29.3} & \text{68.8} & 57.6 & 31.7 & 41.2 & 14.6 & 67.2 & \text{84.1} & \text{28.8} & 39.5 \\
    Multi-Task & 37.5 & 17.4 & 23.8 & 15.5 & 31.7 & 26.9 & 66.9 & 58.1 & 34.8 & 44.5 & 14.2 & 68.1 & 81.0 & 27.6 & 39.1 \\
    \ours & \text{38.4} & 17.6 & 22.7 & 15.3 & 32.1 & 27.4 & 67.4 &59.9 &32.7 & 56.2 & 14.6 & 77.3 & 83.9 & 31.4 & \textbf{41.7}\\
    \midrule
    \datawise & 40.2 & 19.7 & 25.6 & 17.7 & 36.1 & 29.3 & 70.8 & 59.9 & 34.4 & 67.3 & 16.2 & 79.8 & 84.5 & 32.2 & 46.9 \\
    \oracle & 48.6 & 27.0 & 35.2 & 21.5 & 43.1 & 41.1 & 80.3 & 65.1 & 41.1  & 69.1 & 18.2 & 84.1 & 86.4 & 37.1 & 49.9\\
    \bottomrule
    \end{tabular}
\caption
     {\ours performance with eight gates: AR, NF, SF, FI, HO, MS, TR, QU. Avg is the average performance of all datasets.} 
\label{table: router_retriever.expert_8_tr}
\end{table*}

\begin{table*}[t!]
\centering
\fontsize{6.5}{10} \selectfont
    \begin{tabular}{c|ccc|ccccc|cc|cc|c|c|c}
    \toprule
    & \multicolumn{3}{c}{Misc} & \multicolumn{5}{c}{Wiki} & \multicolumn{2}{c}{Bio} & \multicolumn{2}{c}{Science} & \multicolumn{1}{c}{Quora} &\multicolumn{1}{c}{Finance} \\
    \midrule
     & \textbf{AR} & TO & \text{MS} & CL & DB & NQ & FE & \textbf{HO} & \textbf{NF} & TR & SD & \textbf{SF} & \textbf{QU} & \textbf{FI}&Avg   \\
    \midrule
    MSMARCO & 37.2 & \text{18.3} & \text{25.7} & \text{16.0} & 32.7 & \text{29.3} & \text{68.8} & 57.6 & 31.7 & 41.2 & 14.6 & 67.2 & \text{84.1} & 28.8 & 39.5 \\ % 47.5 
    Multi-Task & 35.3 & 17.6 & 21.5 & 15.8 & 31.0 & 25.9 & 66.4 & 55.3 & 32.3 & \text{41.3} & 14.5 & 65.3 & 82.8 & 29.3 & 38.2 \\
    \ours & \text{39.5} & 17.2 & 22.2 & \text{16.0} & \text{32.8} & 27.3 & 68.1 & \text{59.5} & \text{33.4} & \text{44.3} & \text{15.1} & \text{76.0} & 83.6 & \text{30.5} & \textbf{40.4} \\ 
     \midrule
    \datawise & 40.2 & 19.7 & 22.6 & 17.7 & 36.1 & 28.8 & 68.9 & 59.9 & 34.4 & 49.7 &  16.2 & 79.8 & 84.5 & 32.2 &  42.2\\
    \oracle & 48.0 & 26.8 & 32.7 & 21.7 & 40.6 & 39.3 & 80.0 & 65.5 & 38.8 & 56.2 & 21.0 & 85.0 & 89.7 & 39.2 & 48.9 \\
    \bottomrule
    \end{tabular}
\caption
     {\ours performance with six gates: AR, NF, SF, FI, HO, QU. Avg is the average performance of all datasets. The number of total training datasets of \ours, Multi-Task, and MSMARCO-only are the same.} 
\label{table: router_retriever.expert_6}
\end{table*}

\begin{table*}[t!]
\centering
\fontsize{6.5}{10} \selectfont
    \begin{tabular}{c|ccc|ccccc|cc|cc|c|c|cc}
    \toprule
    & \multicolumn{3}{c}{Misc} & \multicolumn{5}{c}{Wiki} & \multicolumn{2}{c}{Bio} & \multicolumn{2}{c}{Science} & \multicolumn{1}{c}{Quora} &\multicolumn{1}{c}{Finance}  \\
    \midrule
     & \textbf{AR} & TO & \textbf{MS} & CL & DB & NQ & FE & \textbf{HO} & \textbf{NF} & TR & SD & \textbf{SF} & \textbf{QU} & \textbf{FI}&Avg \\
    \midrule
    MSMARCO & 37.2 & \text{18.3} & \text{25.7} & 16.0 & 32.7 & \text{29.3} & 68.8 & \text{57.6} & 31.7 & 41.2 & 14.6 & 67.2 & \text{84.1} & 28.8 &  38.5 \\
    Multi-Task & 36.9 & 17.3 & 22.4 & \text{16.3} & 32.9 & 26.7 & 69.0 & 52.1 & 32.9 & 41.4 & 14.5 & 69.4 & 82.0 & 28.9 & 38.8 \\
    \ours & \text{38.6} & 17.7 & 23.0 & 15.2 & \text{33.9} & 27.6 & \text{69.2} & \text{59.9} & \text{33.4} & \text{44.9} & \text{14.8} & \text{77.6} & 83.8 & \text{30.8} & \textbf{40.7} \\
    \midrule
    Best Individual & 40.2 & 19.7 & 25.6 & 17.7 & 36.1 & 29.3 & 70.8 & 59.9 & 34.4 & 49.7 & 16.2 & 79.8 & 84.5 & 32.2 & 42.6 \\
    Oracle &48.5 & 27.2 & 34.5 & 22.3 & 41.5 & 40.8 & 81.0 & 66.6 & 39.0 & 56.4 & 21.2 & 85.4 & 89.9 & 39.6 & 49.6 \\
    \bottomrule
    \end{tabular}
\caption
     {\ours performance with seven gates: AR, NF, SF, FI, HO, QU, MS. Avg is the average performance of all datasets.} 
\label{table: router_retriever.expert_7}
\end{table*}

\subsection{Generalization to other datasets} \label{app: b7}

We observe that \ours demonstrates stable performance not only on datasets with corresponding gates but also on those without them. The performance with different numbers of gates is shown in the following tables: Table~\ref{table: router_retriever.expert_4.ood} (4 gates), Table~\ref{table: router_retriever.expert_5} (5 gates), Table~\ref{table: router_retriever.expert_6} (6 gates), Table~\ref{table: router_retriever.expert_7} (7 gates), and Tables~\ref{table: router_retriever.expert_8_tr} and~\ref{table: router_retriever.expert_8_sd} (8 gates).

When using a similar total number of training datasets (Table~\ref{table: router_retriever.expert_6}), \ours and the MSMARCO-trained model exhibit comparable generalization performance (both at 31.6). However, \ours achieves higher performance on datasets that have corresponding gates (47.5 for MSMARCO-only vs. 49.3 for \ours). As more gates are added, both generalization ability and performance on datasets with corresponding gates tend to improve (Figure~\ref{fig:cases}).

\begin{table*}[t!]
\centering
\fontsize{6.5}{10} \selectfont
    \begin{tabular}{c|cccc}
    \toprule
\textbf{Method}        & \textbf{Training for Routing} & \textbf{Training Datasets for Routing or Experts} & \textbf{Offline Computation} & \textbf{Storage for Pilot Embedding Library} \\ \hline
Single model on MSMARCO                & No                                      & -                                                     & -                           & -                                         \\ \hline
Single model with Multi-task             & No                                      & All $D$ domains                                       & -                           & -                                         \\ \hline
ClassificationHeadRouter    & Yes                                     & All $D$ domains                                       & -                           & -                                         \\ \hline
ExpertClassifierRouter      & Yes                                     & Only the new domain                                   & -                           & -                                         \\ \hline
DatasetRouter         & No                                      & Only the new domain                                   & $T$                         & $D \times T$                              \\ \hline
\textbf{RouterRetriever}          & No                                      & Only the new domain                                   & $D \times T$                & $D \times D$                              \\ 
    \bottomrule
    \end{tabular}
\caption{Comparison of routing methods based on training, computation, and storage requirements. $D$ is the number of domains and $T$ is the number of datasets in the new domain. When new domain is added, ``Training for Routing'' indicates whether the method requires training new router and ``Training Datasets for Routing or Experts'' indicates number of training datasets required for routing or experts. ``Offline computation'' shows how much offline computation is required for routing. ``Storage for pilot embedding library'' indicates how much pilot embeddings are saved in the library for routing.}
\label{tab:efficiency}
\end{table*}

\subsection{Efficiency} \label{app: b8}

Table~\ref{tab:efficiency} summarizes the efficiency details of \ours compared to the baselines.
While \ours requires offline computation to construct the pilot embedding library for routing (as shown in the “Offline Computation” column), it does not necessitate additional training for routing (see “Training for Routing” column). As a result, even though \ours involves offline inference computation, the overall cost remains significantly lower than the expense of training for routing.

DatasetRouter, which also does not require training for routing, incurs less offline computation for routing than \ours. Specifically, while DatasetRouter’s computation scales with $T$, the number of datasets in the new domain, \ours scales with $D \times T$, the number of domains multiplied by number of datasets in the new domain. However, \ours is much storage-efficient where DatasetRouter need $D \times T$ storage and \ours requires $D \times D$ storage. Please note that in most cases $T$ is much larger than $D$. For example, if the initial model was trained with three domains and a new domain with 100 training instances is introduced, $D$ becomes 4, and $T$ is 100. In this case, DatasetRouter performs 100 offline computations, whereas \ours requires 400. However, \ours is more storage-efficient (as detailed in the “Storage for Pilot Embedding Library” column), since DatasetRouter needs to store 100 new embeddings, while \ours only stores 4 embeddings for the new domain. Finally, it is crucial to note that \ours achieves the highest overall performance, as presented in Table~\ref{table: general_perf}.

Regarding computation for training experts when adding a new expert, while the training of the expert itself is necessary, all baselines except MSMARCO require this step. However, the efficiency diverges when considering the column ``Training Datasets for Routing or Experts.'' Both the multi-task and ClassificationHead baselines require retraining from scratch with all datasets, including those from previously trained domains, whenever a new domain is added. In contrast, ExpertClassifierRouter, DatasetRouter, and \ours only require the training dataset for the new domain to train the corresponding expert.

\section{Analysis}

\begin{figure}[t!]
\centering
    \centering
    \begin{minipage}[t]{\linewidth}
        \centering
        \includegraphics[width=0.9\linewidth]{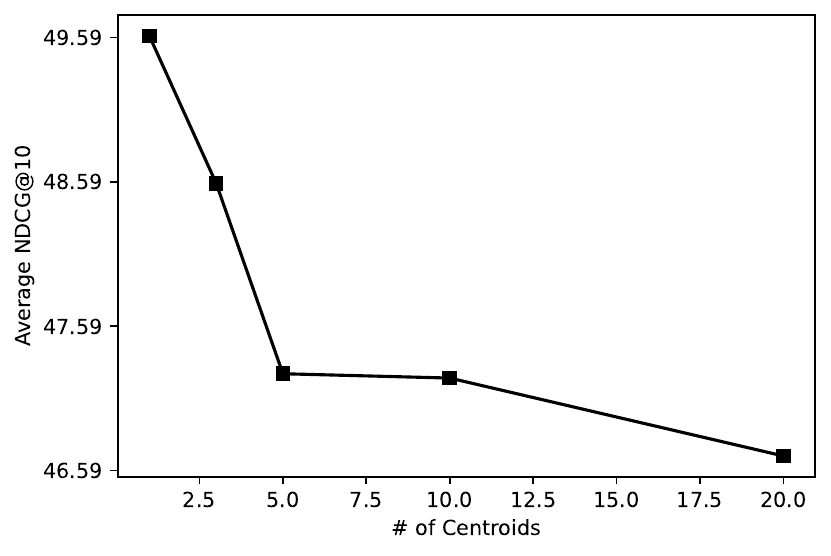}
    \end{minipage}  
        \caption{Average NDCG@10 performance (y-axis) as the number of centroid embeddings from k-means clustering increases. The performance tend to decrease with more pilot embeddings, which suggests that when there are too many pilot embeddings, it tends to distract the performance.} 
        \label{fig:samplenum}
\end{figure}
\subsection{Affect of Number of Pilot Embeddings} \label{app: c1}

We experiment with how the number of pilot embeddings affects performance. In Figure~\ref{fig:samplenum}, we observe that performance tends to degrade as the number of pilot embeddings increases. We hypothesize that this decline is due to the increased number of pilot embeddings becoming distracting, leading to less effective routing decisions.

\begin{table*}[t!]
\centering
\fontsize{6.5}{10} \selectfont
    \begin{tabular}{cc|ccc|ccccc|cc|cc|c|c}
    \toprule
    Domain & Training Data  & AR & TO & MS & CL & DB & NQ & FE & HO & NF & TR & SD & SF & QU & FI \\
    \midrule
    \multirow{2}{*}{Misc}
    & AR   & \textbf{40.2} & 17.0 & 22.4 & 15.7 & 31.0 & 26.2 & 68.9 & 54.2 & 32.8 & 40.3 & 15.5 & 67.8 & 83.7 & 29.6 \\
    & MS  & 37.2 & 18.3 & \textbf{25.7} & 16.0 & 32.7 & \textbf{29.3} & 68.8 & 57.6 & 31.7 & 41.2 & 14.6 & 67.2 & 84.1 & 28.8 \\
    \midrule
    \multirow{1}{*}{Wiki}
    & HO  &  38.5 & \textbf{19.7} & 22.4 & \textbf{17.7} & \textbf{36.1} & 28.8 & 67.8 & \textbf{59.9} & 32.2 & 39.1 & 16.2 & 66.0 & 82.8 & 27.7  \\
    \midrule
    \multirow{2}{*}{Bio}
    & NF & 38.7 & 17.7 & 21.8 & 13.4 & 28.7 & 24.0 & 64.6 & 46.4 & \textbf{34.4} & 42.1 & 15.5 & 66.6 & 82.5 & 27.3 \\
    & TR  & 37.0 & 17.3 & 22.8 & 16.2 & 31.6 & 26.4 & 68.1 & 56.6 & 33.1 & \textbf{67.3} & 15.7 & 68.3 & 83.1 & 29.1  \\
    \midrule
    \multirow{2}{*}{Science}
    & SD &  38.9 & 18.2 & 22.8 & 17.0 & 32.0 & 27.3 & \textbf{70.0} & 57.2 & 33.2 & 39.6 & \textbf{16.3} & 66.7 & 84.3 & 28.4 \\
    &SF & 37.9 & 16.5 & 21.8 & 16.0 & 29.4 & 25.5 & 68.1 & 50.2 & 32.3 & 28.8 & 15.1 & \textbf{79.8} & 83.6 & 25.1 \\
    \midrule
    Quora & QU  & 37.5 & 19.4 & 22.6 & 13.9 & 29.0 & 27.4 & 63.8 & 49.1 & 31.1 & 49.7 & 14.3 & 65.7 & \textbf{84.5} & 28.3 \\
    \midrule
    Finance &  FI  & 35.1 & 18.5 & 22.1 & 15.4 & 29.5 & 25.2 & 64.6 & 46.9 & 32.3 & 42.7 & 15.0 & 64.5 & 83.4 & \textbf{32.2} \\ 
    \bottomrule
    \end{tabular}
\caption
     {Overall performance when evaluating each gate separately.}
\label{table: sep}
\end{table*}

\subsection{Performance of each gates} \label{app: b1}

To analyze the performance trends of each gate, we evaluate them individually without applying any routing techniques in Table~\ref{table: sep}. 
The performance generally shows the highest when the evaluation dataset matches the training dataset of the gate. Additionally, the performance gap between matching and non-matching datasets is larger for domain-specific datasets (NF, TR, SD, SF, QU, FI). In contrast, gates trained on general-domain datasets (AR, MS, HO) tend to perform well across a broader range of datasets.

\begin{figure}[t!]
\centering
    \centering
    \begin{minipage}[t]{\linewidth}
        \centering
        \includegraphics[width=0.85\linewidth]{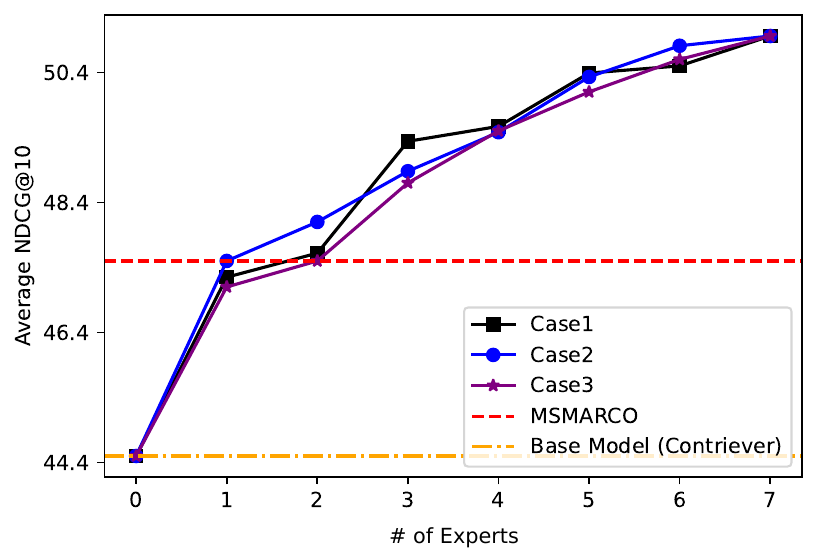}
    \end{minipage}  
        \caption{For each evaluation dataset (y-axis), the rate at which gate the router chooses (x-axis). We could see that the trend generalizes with different combination of gates. Case1 is in order of AR, FI, SF, NF, HO, QU, and MS. Case2 is in reverse order of MS, QU, HO, NF, SF, FI, and AR. Case 3 is in order of SF, NF, HO, QU, AR, MS, and FI.} 
        \label{fig:cases}
\end{figure}

\subsection{Impact of Number of Gates} \label{app: c2}
To investigate the impact of the number of gates, we randomly shuffle the gate order and experiment with how adding gates tend to affect performance.
The order of gates added in Figure 4 and Figure 5 is AR, FI, SF, NF, HO, QU, and MS.
We tried various other combinations and could see that the findings are stabilized (Figure~\ref{fig:cases}): (1) performance tend to increase with more gates added and (2) the improvement rate tends to be higher when adding gates initially and as the number of gates grows, the rate of increase diminishes.

\begin{table*}[t!]
\centering
\fontsize{6.5}{10} \selectfont
    \begin{tabular}{c|cc|c|cc|cc|c|c|ccc}
    \toprule
    & \multicolumn{2}{c}{Misc} & \multicolumn{1}{c}{Wiki} & \multicolumn{2}{c}{Bio} & \multicolumn{2}{c}{Science} & \multicolumn{1}{c}{Quora} &\multicolumn{1}{c}{Finance} \\
    \midrule
     & \text{AR} & \text{MS} & \text{HO} & \text{NF} & \text{TR} & \text{SD} & \text{SF} & \text{QU} & \text{FI} & Avg \\
    \midrule
    \ours & \text{39.5}  & 22.2 & \text{59.5} & \text{33.4} & \text{44.3} & \text{15.1} & \text{76.0} & 83.6 & \text{30.5} & 44.9 \\ 
    \ours~(+ TR) & 38.4  & 22.7 &59.9 & 33.3 & \textbf{56.2} & 14.6 & 77.3 & 83.9 & \text{31.4} & 46.4 \\ 
    \ours~(+ SD) & 38.8 & 22.9 & 59.8 & 32.7 & 44.7 & \textbf{16.1} & 76.9 & \text{84.1} & 30.3 & 45.1 \\ 
    \bottomrule
    \end{tabular}
\caption
     {\ours performance when adding gates within same domain. The performance tend to improve for the dataset, but for the rest the difference tend to be minor.} 
\label{table: router_retriever.multiple.id}
\end{table*}

\subsection{Routing Mechanism Error Analysis} \label{app: b6}

Figure 7 illustrates the rate at which each router selects a gate, while Figure 6 shows the rate at which each gate tends to deliver high performance for the dataset. The discrepancy between these two heatmaps highlights the gap between \ours and the oracle performance.
For ArguAna, the maximum gate distribution is evenly spread, and the routing tends to follow this distribution closely.  
For Quora, while the maximum gate rate is high overall, the routing often favors the HotpotQA gate in many cases.  
For MSMARCO, the gate trained on MSMARCO generally shows high performance, but the routing technique tends to distribute selections across different gates.  
For HotpotQA, selecting the HotpotQA gate most frequently results in the highest performance, with MSMARCO being the next best option. The routing technique tends to reflect this pattern.  
For SciFact, choosing the SciFact gate is crucial in both cases.  
For NFCorpus, selecting the NFCorpus gate is important, yet the routing technique often opts for the ArguAna gate in many instances.  
For FiQA-2018, the best performance is achieved by selecting the FiQA-2018 gate, and the routing technique successfully identifies this gate most of the time.

We specifically investigated why NFCorpus often fails to select the NFCorpus gate and instead tends to choose the ArguAna gate. Upon examining the representative embeddings for ArguAna, we found that many of them are confused with ArguAna embeddings that were extracted from the NFCorpus dataset. These instances originally belong to NFCorpus but show the highest performance with the ArguAna gate, leading to their labeling as ArguAna. This suggests that instead of completely removing information about the original dataset, incorporating a weighting factor between the two could further improve performance.

\end{document}